\documentclass{mn2e}

\usepackage{amsmath}
\usepackage{amssymb}
\usepackage{graphicx,epsfig}

 \font\sevenrm=cmr7 scaled 1000

\def\gsim{\;\lower4pt\hbox{${\buildrel\displaystyle >\over\sim}$}\;}
\def\lsim{\;\lower4pt\hbox{${\buildrel\displaystyle <\over\sim}$}\;}
\def\grls{\;\lower4pt\hbox{${\buildrel\displaystyle >\over <}$}\;}

\title[Circumstellar Disc Dispersal]
{Dispersal of Gaseous Circumstellar Discs around High-Mass Stars }

\author[Y. Shen \& Y.-Q. Lou ]
{Yue Shen$^{1,4}$ and Yu-Qing Lou$^{1,2,3}$ \\
$^1$Physics Department and Tsinghua Center for
Astrophysics (THCA), Tsinghua University, Beijing 100084, China;\\
louyq@tsinghua.edu.cn; lou@oddjob.uchicago.edu;
yshen@astro.princeton.edu \\
$^2$Department of Astronomy and Astrophysics, The University
of Chicago, 5640 South Ellis Avenue, Chicago, IL 60637 USA;\\
$^3$National Astronomical Observatories of China, Chinese
Academy of Sciences, A20, Datun Road, Beijing 100012, China;\\
$^4$Department of Astrophysical Sciences, Peyton Hall,
Princeton University, Princeton, NJ 08544 USA. }
\date{Accepted 2004... Received 2003...;
in original form 2003}\date{Accepted .
      Received ;
      in original form }
\begin{document}
\maketitle

\begin{abstract}
We study the dispersal of a gaseous disc surrounding a central
high-mass stellar core once this circumstellar disc becomes
fully ionized. If the stellar and surrounding EUV and X-ray
radiations are so strong as to rapidly heat up and ionize the
entire circumstellar disc as further facilitated by disc
magnetohydrodynamic (MHD) turbulence, a shock can be driven to
travel outward in the fully ionized disc, behind which the disc
expands and thins. For an extremely massive and powerful stellar
core, the ionized gas pressure overwhelms the centrifugal and
gravitational forces in the disc. In this limit, we construct
self-similar shock solutions for such an expansion and depletion
phase. As a significant amount of circumstellar gas being removed,
the relic disc becomes vulnerable to strong stellar winds and
fragments into clumps.
%which may further coagulate to form planets. The
%fast moving shock may also trigger planet formation.
We speculate that disc disappearance happens rapidly, perhaps on
a timescale of $\sim 10^3-10^4\hbox{ yr}$ once the disc becomes
entirely ionized sometime after the onset of thermal nuclear
burning in a high-mass stellar core.
\end{abstract}

\begin{keywords}
circumstellar matter -- hydrodynamics --
%planetary systems: formation --
plasmas -- shock waves -- stars: early
type -- stars: winds, outflows
\end{keywords}

\section{INTRODUCTION}

%Young stars are often observed to have circumstellar discs, which
%are the byproducts through star formation processes (Shu et al. 1987;
%Larson 2003). During the pre-main-sequence accretion phase, the
%central stellar object acquires mass via the rotating circumstellar
%disc. After the star reaches the ZAMS and ignites core nuclear
%reaction, it no longer accretes from the disc. Instead, the UV
%radiation and early strong stellar winds cause the remained
%circumstellar disc to expand and deplete.

There are ample evidence that young stars are often accompanied by
circumstellar discs, as part of star formation processes. Discs
are most frequently detected for low-mass T Tauri protostars.
%There are some new evidence that discs may be also acquired
%from a dense interstellar medium (e.g. Abt 2004).
The probability of detecting a circumstellar disc decreases with
an increasing stellar mass. Natta et al. (2000) concluded that a
lack of discs around Herbig Be stars may result from their faster
evolution. Fuente et al. (2003) revealed the first evidence of
dusty discs around Herbig Be stars, yet the disc-to-star mass
ratio $M_d/M_*$ is at least an order of magnitude lower than those
of Herbig Ae and T Tauri stars, suggesting that discs may evolve
much more rapidly around very massive stars. Suppose that high-mass
stars form similarly as low- or intermediate-mass stars in the sense
of involving collapse (e.g., Lou 1996; Lou \& Shen 2004; Bian \& Lou
2005; Yu \& Lou 2005; Yu, Lou, Bian \& Wu 2006; Lou \& Gao 2006) and
accretion (e.g., McKee \& Tan 2002; Blum et al. 2004), then a
circumstellar disc should form during the early evolution phase.
Observational evidence for accretion and massive discs during
high-mass star forming stages
were indeed reported (e.g., Sandell et al. 2003; Pestalozzi et al.
2004). While massive stars of late $O$ or early $B$ types may have
disc signatures, more massive stars lack clear evidence for discs
(e.g., Blum et al. 2004). Unless a massive star forms through
other processes without involving a circumstellar disc, the key
issue is to understand the mechanism for a fast and efficient
dispersal (primarily hydrogen gas) of circumstellar discs around
the most massive stars.

Circumstellar discs are important because they are sites where
a companion star or planets may form. For a massive star, a
considerable mass fraction of its dense circumstellar disc
appears to be rapidly removed.
%before planets form. The timescale of disc dispersal is
%thus a crucial parameter relevant to the planet formation.
The currently estimated lifetime of circumstellar discs for
low-mass stars is of the order of several million years (e.g.,
Strom et al. 1993; Brice\~{n}o et al. 2001). For intermediate- and
high-mass stars, the timescale for the existence of circumstellar
discs can be much shorter because of faster evolution and stronger
radiative power output from central stars (e.g., Natta et al. 2000).
Theoretically, the dispersal or destruction of a circumstellar
disc may be modelled through viscous disc accretion onto a central
star, photoevaporation by either stellar or external UV radiations,
interactions with other stars or consumption by planet formation, or
any combinations of these processes (e.g., Hollenbach et al. 2000).
The photoevaporation and accretion models have been widely applied
to disc dispersals of low-mass T Tauri stars (e.g., Johnstone et al.
1998; St\"orzer \& Hollenbach 1999; Clarke et al. 2001).
%During the pre-main-sequence phase, the circumstellar disc
%may be massive since the star must accrete a considerable
%portion of its final mass via the disc (Sandell et al. 2003).
%Finally, the accretion ceases and an optically thick disc
%remains outside the star. Shortly after the ignition of the
%star, the disc is rapidly ionized and expands (perhaps also
%under the influence of early stellar winds). During this
%expansion phase, a considerable amount of the disc mass is
%transported outward. This may explain the presence of
%appreciable amounts of ionized gas at intermediate radii in
%ultracompact H {\sevenrm II} regions (Shu et al. 2002).

We here advance a plausible physical scenario for the dispersal of
circumstellar discs around most massive stars. By forming a
massive protostellar core, the accretion gradually ceases and a
massive (perhaps $\gtrsim 0.1-0.3 M_\odot$) centrifugally
supported H {\!\sevenrm I} disc\footnote{Discs around massive
stars may be even more massive initially (see Sandell et al.
2003).} forms around the core. Near the main-sequence
%\footnote{This switch time may correspond to the start of the
%main-sequence or to a phase of the pre-main-sequence evolution.}
when the stellar core ignites thermal nuclear reactions to produce a
profuse amount of extreme-ultraviolet (EUV) photons, an ionization
front (IF) travels rapidly outward to ionize the entire H
{\!\sevenrm I} disc to a high temperature $\gsim 10^4$ K (e.g.,
Osterbrock 1989). Meanwhile, X-ray emissions associated with
protostellar and disc magnetic activities also ionize gas to
$\sim 10^6-10^7$ K within a few thousand AUs (e.g., Glassgold
et al. 2000). For a high-mass star of $M_*\sim 10M_{\odot}$, the
stellar luminosity alone may not fully ionize a very dense
circumstellar disc. However, a circumstellar disc may be fully
ionized if (a) the central stellar mass $M_*$ is much greater
than $10M_{\odot}$ or the Lyman flux is considerably higher;
(b) the host star is born in a cluster, where its circumstellar
disc is exposed to intense external X-ray and UV radiations;
(c) a circumstellar disc has been diluted significantly via other
processes such as photoevaporation (e.g., Hollenbach et al. 1994);
(d) magnetorotational instabilities
(e.g., Balbus \& Hawley 1998; Balbus 2003) in the weak magnetic
field regime sustain a level of MHD turbulence to incessantly mix
ionized and neutral components and to effectively enhance
ionization. To estimate, we assume a diluted neutral H {\!\sevenrm
I} disc\footnote{Disc radii $R$ of high-mass stars are expected to
be larger than those of T Tauri or Herbig Ae/Be stars, which
extend from tens to hundreds of AUs (e.g., McCaughrean 1997; Meyer
\& Beckwith 2000). We take $R\sim 10^{3}\hbox{ AUs}$ (e.g.,
Pestalozzi et al. 2004).} with a surface mass density
$\Sigma(r)=\Sigma_d(r/R)^{-1/2}$. For $M_d\sim 0.02M_{\odot}$, we
derive $\Sigma_d=3M_d/(4\pi R^2)\sim 0.042\hbox{ g cm}^{-2}$ such
that the H {\!\sevenrm I} column density normal to the disc is
$N_H(r)\sim 2.5\times 10^{22}(r/R)^{-1/2}\hbox{ cm}^{-2}$. The
Lyman flux incident on the disc will penetrate a column
$N_e(r)\lesssim 1.8\times 10^{22}\Phi_{49}^{1/2}(r/R)^{-1/2}
\hbox{ cm}^{-2}$ (see Appendix A of Hollenbach et al. 1994). Thus
an H {\!\sevenrm I} disc may be fully ionized\footnote{Around
massive stars, dust grains are effectively evacuated to larger
radii ($\sim 0.1\hbox{ pc}$) due to sublimation, stellar winds and
radiation pressure (e.g., Chini et al. 1987; Churchwell 2002).
When dust attenuation is included, the required relic disc mass
will be smaller. In the approximation of Franco et al. (1990),
the incident Lyman flux $\Phi_{49}$ is reduced by a factor
$e^{-\tau_d}$ where $\tau_d$ is the optical depth at the boundary
of H {\!\sevenrm II} regions. For $\tau_d\gsim 1$, the disc mass
that can be fully ionized is reduced by a factor less than $0.6$.}
for $\Phi_{49}\gsim 2$, which is readily satisfied for early $O$
stars (e.g., Panagia 1973). For more massive stars with
$\Phi_{49}\sim 10-100$, a denser circumstellar disc of mass $M_d$
even higher than $\sim 0.2M_{\odot}$ might be ionized completely
as facilitated by X-ray fluxes and disc MHD turbulence. If
submerged in an external X-ray and EUV radiation field from nearby
massive stars within a local star cluster (which is the usual case),
a complete ionization of a circumstellar disc becomes more likely.
%However, the observed neutral disc is often optically thick.
%For typical values, a circumstellar neutral disc with radius
%$R\sim 1000\hbox{ AU}$, mass $M_d\sim 0.2M_{\odot}$ and scale
%height $H\sim 0.1R$ has a average $\hbox{H}_2$ density of
%$n_{H_2}\sim 10^8\hbox{ cm}^{-3}$. The Str\"{o}mgren radius
%is $r_s\sim 40\hbox{ AU}$.} $\sim 10^4-10^5\hbox{ K}$.
%{\bf A temperature of $10^4$ or $10^5$ K or even higher?
A hot H {\!\sevenrm II} disc inevitably expands as a result
of the overwhelming pressure gradient. Typically, the gas
density is higher in the inner part than in the outer part
of a disc, so does the pressure gradient. Hence, the inner
disc expands faster than the outer disc does and consequently
a shock develops in the disc to travel radially outward. The
disc portion behind the shock will continue this expansion to
reduce the disc surface mass density.

The main thrust of this Letter is to construct a gas dynamic model
for a such disc expansion phase after a circumstellar disc has
been ionized and to estimate disc disappearance timescales using
available data. While our results are preferably applicable to
high-mass stars ($\gsim 10M_{\odot}$), they are also relevant for
discs around low- or intermediate-mass stars in binaries where the
disc of the secondary may be fully ionized by the X-ray and EUV
field of the more massive primary.

\section{DISC MODEL FORMULATION}

We start with a specific case of a neutral H {\sevenrm I} disc
carrying a surface mass density $\Sigma\propto r^{-1}$, and
then extend the results to different power-law indices of
$\Sigma$ profiles (Shen \& Lou 2004a; Shen, Liu \& Lou 2005;
Lou \& Bai 2006; Wu \& Lou 2006) in the limit when the thermal
pressure dominates other forces in the shock evolution of an
H {\sevenrm II} disc. The stellar gravity has weaker effects
over distances larger than $\tilde{r}\sim GM_*/(2a^2)$ where
$a$ is the sound speed of the H {\sevenrm II} gas. For
$M_*\sim 10M_{\odot}$ and $a\sim 10\hbox{ km s}^{-1}$, we have
$\tilde{r}\sim 40\hbox{ AU}$ and may neglect the stellar gravity
on the main portion of the disc with a typical radius
$R\sim 1000\hbox{ AU}$.

We presume a thin neutral H {\!\sevenrm I} circumstellar disc
around a protostar in a rotational equilibrium initially with a
singular isothermal disc (SID) profile, namely
%with a flat rotation curve $V_0=a_0D$ and axisymmetry,
%\footnote{Why ignore the gravity of the star? The gravity of
%the central star causes a gap between the star and the disc?
%For typical $D=2$, $r=100AU$, $a_0=0.2 kms^{-1}$ the disc mass
%is $M=0.02 M_{\odot}$, consistent with observation. Also there
%are increasingly evidence that the circumstellar discs are of surface mass
%density proportional to $r^{-1}$. Saigo \& Hanawa (1998) also
%showed a centrifugally supported SID can form during the
%accretion phase of star formation. We here choose the SID
%configuration for its simplicity }
\begin{equation}\label{SID}
\Sigma(r)=\frac{a_0^2(1+D^2)}{2\pi Gr}\ ,\ \
M(r)=\frac{a_0^2(1+D^2)r}{G}\ ,\ \ j(r)=a_0Dr\ ,
\end{equation}
where in cylindrical coordinates $(r,\theta,z)$, $\Sigma(r)$, $M(r)$
and $j(r)$ are the surface mass density, the enclosed mass inside
radius $r$ and the $\hat z-$component specific angular momentum,
respectively; $G$ is the gravitational constant and $D$ is the
rotational Mach number defined by $D\equiv V_0/a_0$ with $V_0$
and $a_0$ being the rotation speed\footnote{We take on a flat
rotation curve instead of the usual Keplerian rotation curve.
%(Korner's paper).
As can be seen later, the centrifugal force can be omitted
in the limit when the pressure dominates and thus the choice
of different rotation curves does not really matter.}
and the isothermal sound speed of the H {\!\sevenrm I} gas.
%The radial flow speed is $U_0=0$.

At a certain initial time $t=0$, thermal nuclear reactions start
in the stellar core and the resulting radiation rapidly ionizes
the circumstellar disc in a short time to reach an isothermal
state with a faster sound speed $a$. To keep the initial SID
profile unchanged would require
\begin{equation}\label{density}
\begin{split}
&\Sigma(r,\ 0^+)=\frac{\epsilon^2 a^2(1+D^2)}{2\pi Gr}\ ,
\qquad j(r,\ 0^+)=\epsilon aDr\ ,\\
&M(r,0^+)=\frac{\epsilon^2a^2(1+D^2)r}{G}\ ,
\end{split}
\end{equation}
where $\epsilon\equiv a_0/a\le 1$ is the square root of the
temperature ratio. Now the pressure and centrifugal forces
overwhelm the gravitational force and a circumstellar disc expands
inevitably. As the pressure gradient is steeper in the inner disc,
a shock will emerge to travel outward. This is a close analog of
the so-called `champagne flows' in H {\sevenrm II} clouds
surrounding OB stars (Tenorio-Tagle 1979; Franco et al. 1990).
Since the latter may evolve towards a self-similar phase (Shu et
al. 2002), it is suggestive of a similar description for a disc
expansion with a shock.
%A complete classification of various self-similar solutions in a
%thin disc may be found elsewhere (Shen \& Lou 2004 in preparation).
Here, we focus on axisymmetric similarity shocks relevant
to the disc dispersal scenario.

Under the axisymmetry, we use the following nonlinear
equations to describe disc dynamics in cylindrical geometry.
\begin{equation}\label{mass}
\frac{\partial\Sigma}{\partial t}+\frac{1}{r}
\frac{\partial}{\partial r}(r\Sigma u)=0\ ,
\end{equation}
\begin{equation}\label{rmomentum}
\frac{\partial u}{\partial t}+u\frac{\partial u}{\partial r}
-\frac{j^2}{r^3}=-\frac{1}{\Sigma}\frac{\partial\Pi}{\partial r}
-\frac{\partial\Phi}{\partial r}\ ,
\end{equation}
\begin{equation}\label{phimomentum}
\frac{\partial j}{\partial t}+u\frac{\partial j}{\partial r}=0\ ,
%\end{equation}
%\begin{equation}\label{encmass}
\qquad\qquad
\frac{\partial M}{\partial t}+u\frac{\partial M}{\partial r}=0\ ,
\end{equation}
where $u$ is the radial speed, $\Phi$ is the gravitational
potential, $\Pi$ is the two-dimensional thermal gas
pressure, $M(r,t)\equiv\int_0^r\Sigma(r^{\prime},t)2\pi
r^{\prime}dr^{\prime}$ is the enclosed mass within radius $r$.
The Poisson integral is
\begin{equation}\label{Poisson}
\Phi(r, t)=\int\int\frac{-G\Sigma(r^{\prime}, t)
r^{\prime}\mathrm{d}r^{\prime}
\mathrm{d}\theta}{(r^{\prime 2}+r^2
-2r^{\prime}r\cos\theta)^{1/2}}\ .
\end{equation}
%where $G$ is the gravitational constant.
%
We introduce the following similarity transformation
\begin{equation}\label{transform1}
\begin{split}
&x\equiv\frac{r}{at}\ ,\qquad
\Sigma(r,t)\equiv\frac{a\alpha(x)}{2\pi Gt}\ ,
\qquad
u(r,t)\equiv av(x)\ ,\\
&M(r,t)\equiv\frac{a^3t}{G}m(x)\ ,
\qquad\qquad j(r,t)\equiv\beta a^2tm(x)\
\end{split}
\end{equation}
in the two-dimensional isothermal self-gravitating nonlinear
ideal gas equations (\ref{mass})$-$(\ref{Poisson}), where
$\alpha(x)$, $v(x)$ and $m(x)$ are the reduced surface mass
density, radial flow speed and enclosed mass, all being
dimensionless functions of $x$. Parameter $\beta\equiv D/
[\epsilon (1+D^2)]$ is determined from the equilibrium. The
last one in transformation (\ref{transform1}) holds as the
ratio $j(r)/M(r)$ remains constant since $t=0$ by the angular
momentum conservation (Li \& Shu 1997; Saigo \& Hanawa 1998).

In the so-called monopole approximation,\footnote{A multipole
treatment can be found elsewhere (Li \& Shu 1997). This
monopole approximation does not affect our results much
because when $\epsilon \ll 1$,
%which is relevant for the present problem,
the self-gravity becomes negligible
%compared with the thermal pressure
(Shu et al. 2002; Shen \& Lou 2004b).} the similarity
ordinary differential equations (ODEs) become
\begin{equation}\label{ode1}
\frac{[(x-v)^2-1]}{(x-v)}\frac{dv}{dx}=\frac
{\alpha (x-v)[1-\beta^2\alpha(x-v)]-1}{x}\ ,
\end{equation}
\begin{equation}\label{ode2}
\frac{[(x-v)^2-1]}{(x-v)\alpha}\frac{d\alpha}{dx}
=\frac{\alpha[1-\beta^2\alpha(x-v)]-(x-v)}{x}\ ,
\end{equation}
%\begin{equation}\label{ode3}
%m=x(x-v)\alpha\ ,
%\end{equation}
which, together with $m=x(x-v)\alpha$, are to
be solved for specified `asymptotic conditions'.

The leading asymptotic solutions at $x\rightarrow +\infty$
for either $t\rightarrow 0^+$ or $r\rightarrow\infty$ are
\begin{equation}\label{asyinf1}
\begin{split}
v\rightarrow V+\frac{1-A+\beta^2A^2}{x}\ ,\quad \ \
%\frac{V(1-\beta^2A^2)}{2x^2}
%+\frac{2V^2+(A-1-\beta^2A^2)(A-4)}{6x^3}\ ,\\
\alpha\rightarrow\frac{A}{x}\ ,\quad \ \
%+\frac{A(1-A+\beta^2A^2)}{2x^3}\
%, \qquad\qquad
m\rightarrow Ax\ ,
\end{split}
\end{equation}
where $V$ and $A$ are two constants and higher-order
terms can also be readily derived. With an initial
SID, boundary conditions for $x\rightarrow +\infty$
are $V=0$ and $A=\epsilon^2(1+D^2)$.

There exist two classes of asymptotic solutions at
$x\rightarrow 0$ for either $r\rightarrow 0$ or
$t\rightarrow\infty$; we focus on the class of
regular solutions with leading behaviours of
\begin{equation}\label{asym0regular}
\begin{split}
v(x)\rightarrow x/2\ ,\qquad
%-\frac{B}{12}x^2+\frac{1}{32}
%\bigg[1+\bigg(\frac{1}{3}+\beta^2\bigg)B^2\bigg]x^3+...\ ,\\
\alpha(x)\rightarrow B\ ,\qquad
%-\frac{B^2}{2}x+\frac{B}{8}\bigg[1+\bigg(\frac{5}{3}
%+\beta^2\bigg)B^2\bigg]x^2+...\ ,\\
m(x)\rightarrow Bx^2/2\
%-\frac{B^2}{6}x^3+...\
\end{split}
\end{equation}
%These higher-order terms were commented out above for the
%brevity of an ApJL paper. They were actually used in numerical
%computations of solving ordinary differential equations.
as $x\rightarrow 0$, where $B$ is an integration constant.

The two isothermal shock jump conditions are
\begin{equation}\label{shockcond}
(v_d-x_s)(v_u-x_s)=1\ ,\qquad\qquad
\alpha_u=(v_d-x_s)^2\alpha_d\ ,
\end{equation}
where subscript $d$ ($u$) denotes the downstream (upstream) of a
shock and $x_s$ is the `shock location' in the similarity variable
$x$ with a shock speed $ax_s$ (Shen \& Lou 2004b; Bian \& Lou 2005;
Yu et al. 2006).

To solve ODEs (\ref{ode1}) and (\ref{ode2})
numerically, we specify $\beta$ leading to
%In our numerical procedure we choose$0\le\beta<0.5?$.
a relation for $D$ and $\epsilon$
by $\beta\equiv D/[\epsilon (1+D^2)]$.
%
%To numerically solve ODEs (\ref{ode1})-(\ref{ode3}) one needs to
%know the value of $\beta\equiv D/[\epsilon (1+D^2)]$. For the
%initial equilibrium neutral SID (\ref{SID}), the rotational Mach
%number $D$ is in the range $0.9320<D^2<5.410$ so that the SID is
%stable against axisymmetric perturbations (Lemos et al. 1991; Shu
%et al. 2000). For typical values we fix $D=2$. The value of sound
%speed ratio $\epsilon$ is adjustable. For reasonable values we
%take $\epsilon=0.1$, i.e., the ionized disc has a sound speed
%10 times that of the neutral disc.
%
A shock solution is constructed as follows:
(i) Integrate ODEs (\ref{ode1}) and (\ref{ode2}) from the
origin using solution (\ref{asym0regular}) for various $B$;
(ii) Impose shock condition (\ref{shockcond}) at each
integration step to cross the sonic critical line
$x-v=1$ and continue for larger $x$;
(iii) Pick out the one that matches condition (\ref{asyinf1})
as $x\rightarrow\infty$ with $V=0$. By matching a proper $A$
value, this last step leads to another relation of $D$ and
$\epsilon$ by $A=\epsilon^2(1+D^2)$. Both $D$ and $\epsilon$
are then determined accordingly.

%Now a prospective shock solution can be found by the following
%procedure: 1. Integrate ODEs (\ref{ode1}) and (\ref{ode2}) from
%sufficiently large $x$ backward by using asymptotic solutions
%(\ref{asyinf1}) with $V=0$ and $A=\epsilon^2(1+D^2)$; 2. Impose
%shock jump condition (\ref{shockcond}) at integration steps near
%the sonic critical line $x-v=1$ to cross it and continue
%integrating further in; 3. At the same time, integrate ODEs
%(\ref{ode1}) and (\ref{ode2}) from the origin by using asymptotic
%solutions (\ref{asym0regular}) with various $B$ outward; 4. Match
%the inwardly and outwardly integrations at fixed meeting point
%$x_F$ in the $v$ versus $\alpha$ phase diagram until we find a
%match (Lou \& Shen 2004).

\begin{table}
\center\caption{Parameters for numerical solutions}\label{tab:parameters}
\begin{tabular}{cccccc}\hline
     $B$  & $x_s$& $A$                   &$\beta$& $D$    &$\epsilon$\\
\hline
 0.1      & 1.92 & 0.136                 & 0     & 0      &0.369 \\
          & 1.93 & 0.138                 & 0.5   & 0.189  &0.365 \\
          & 1.96 & 0.144                 & 2.0   & 1.166  &0.247 \\
0.01      & 1.99 & 0.0155                & 0     & 0      &0.124 \\
          & 2.00 & 0.0158                & 2.0   & 0.260  &0.122 \\
          & 2.00 & 0.0158                & 8.0   &--      &--    \\
$10^{-3}$ & 2.00 & 1.58$\times 10^{-3}$  & 0     & 0      &0.0397\\
          & 2.00 & 1.58$\times 10^{-3}$  & 8.0   & 0.335  &0.0377\\
          & 2.00 & 1.58$\times 10^{-3}$  & 32    &--      &--    \\
$10^{-4}$ & 2.00 & 1.58$\times 10^{-4}$  & 0     & 0      &0.0126\\
          & 2.00 & 1.58$\times 10^{-4}$  & 32    & 0.439  &0.0115\\
          & 2.00 & 1.58$\times 10^{-4}$  & 128   &--      &--    \\
\hline
\end{tabular}

The three rows without values for $D$ and $\epsilon$ are
unphysical cases.
%The three rows with two -- are unphysical cases.
\end{table}

Remarkably, it is found that in the limit $B\rightarrow 0$
for a disc heated to high temperatures,
%{\bf what is the meaning of $B$?}{\it $B$ is decided
%by $A$, and $A$ is decided by $\epsilon$, so $B$
%stands for the extent to which the disc is heated.}
there exists an invariant shock solution not affected by
$\beta$. The $A$ value is uniquely determined by $B$ with
$A\rightarrow 0$ as $B\rightarrow 0$, while the ratio $A/B$
approaches a constant $\sim 1.58$ (Table 1). Physical solutions
require $A\beta^2<1$. This bears a strong resemblance to the
spherical similarity solutions of `champagne flows' (Shu et al.
2002). This invariant form describes a situation when the
thermal pressure overwhelms in the disc expansion with negligible
gravity and centrifugal forces. The limit of
$x_s=2.00$ gives a shock speed twice the ionized sound speed
$a$. The downstream (post-shock) gas becomes more uniformly
distributed and a linear expansion emerges from small $x$.

Following the above procedure by ignoring gravity and centrifugal
forces, one can extend these results to a power-law surface mass
density $\Sigma(r)\propto r^{-n}$ where
%$K$ is a dimensional constant and
$0<n\le2$ (Shen et al.
%, Liu \& Lou
2005). Similar to the $n=1$ case, the downstream behind the
shock approaches a uniform density yet with the enclosed mass
$\propto t^{-n}$ and the flow speed becomes linear in radii,
viz., $v\rightarrow nx/2$ as $x\rightarrow 0$.

\section{CIRCUMSTELLAR DISCS}

We use our self-similar invariant shock solutions to estimate a
circumstellar  disc dispersal timescale. To elucidate the concept,
we introduce the following fiducial solution using the SID model.
For a centrifugally supported neutral equilibrium SID,
%{\bf SID involves pressure and self gravity}
we may choose $D=2$ for a supersonic rotation against
self-gravity\footnote{$D$ should be $\gsim 1$, but may
not be too large. Our basic results derived here are
not very sensitive to $D$ variations.}.
%{\bf Any particular reason or plausible
%argument for choosing $D=2$?}
The Lyman ionization may lead to a $a_{0}/a$ ratio of $\epsilon=0.02$
(e.g., $a_0\sim 0.2\hbox{ km s}^{-1}$ and $a\sim 10\hbox{ km s}^{-1}$).
We then have $A=2\times 10^{-3}$ which gives a $B=1.26\times 10^{-3}$
and a $\beta=20$. Yet this solution remains essentially the same as
that of $\beta=0$ and $B=1.26\times 10^{-3}$. That is, there exists
an invariant shock solution (Fig. 1) not affected by $\beta$ in
the limit of $B\rightarrow 0$ with a shock at $x_s=2.00$.
%This invariant shock is displayed in Fig. 1.
%and the relevant values of reduced quantities $v(x)$, $\alpha(x)$,
%$m(x)$ downstream (post-shock) are tabulated in Table 2.

\begin{figure}
\mbox{\epsfig{figure=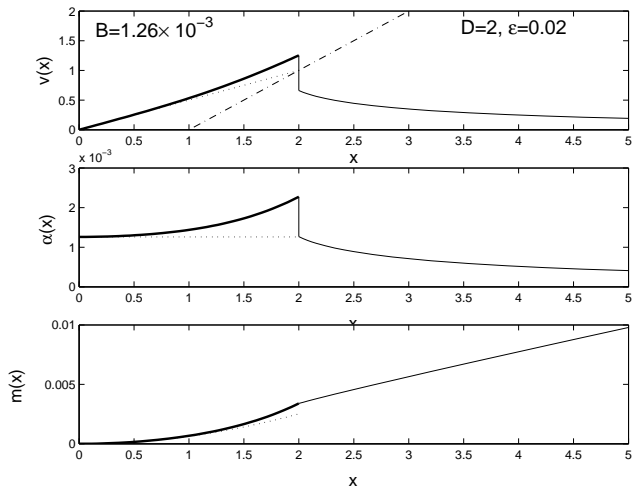,width=\linewidth,clip=}}
\caption{A shock solution for $v(x)$, $\alpha(x)$ and $m(x)$
versus $x$. The dash-dotted line in the top panel is the sonic
critical line $x-v=1$. The reduced radial speed $v(x)$ and
density $\alpha(x)$ encounter discontinuities at shock
location $x_s=2.00$, while the reduced enclosed mass $m(x)$
is continuous by the mass conservation (\ref{mass}) and
%(\ref{ode3})
(\ref{shockcond}). Downstream (post-shock) and upstream
(pre-shock) portions are shown in heavy and light solid
curves; the dotted curves denote the leading-order terms of
asymptotic solutions (\ref{asym0regular}).}\label{fig:example}
\end{figure}

Since a similarity solution does not carry characteristic
timescales or lengths, one needs a starting point to apply the
solution in proper spatial and temporal domains. For example, at a
particular radius $R\sim 1000\hbox{ AU}$, i.e., the ionized disc
radius after some time of expansion, the corresponding similarity
variable is $x_c=R/(at_c)$ where $t_c$ is an input timescale that
a self-similar evolution has lapsed and marks the starting epoch
for the subsequent evolution. Here, $t=t_c$ is a calibration
point. We focus on $R$ and presume that for $t\ge t_c$, all
materials leaving $R$ would be regarded as lost from a
circumstellar disc.
%In this manner, the disc thins as time goes on.
The key issue of concern is
%The key issue of interest is the disc mass depletion,
%characterized by
the evolution of the enclosed disc mass $M(R,\ t)$ for
$t\gsim t_c$ where $t_c$ is a starting time when the mass
and radius of a disc are specified. The information may
be extracted from the bottom panel of Fig. 1, i.e., the
downstream portion of $m(x)$ in heavy solid curve.
%This part of solution is fitted by the following quadratic
%expression:
%\begin{equation}\label{fit}
%m(x)= 0.00025x^{2} - 8.3\times 10^{-5}x + 6\times 10^{-6}\ .
%\end{equation}
%This leads to the evolution of $M(R,t)$ for $t\ge t_0$ as
%\begin{equation}
%M(R,\ t)=\frac{a^3t}{G}\bigg[0.00025\bigg(\frac{2t_0}{t}\bigg)^{2}
%- 8.3\times 10^{-5}\bigg(\frac{2t_0}{t}\bigg) + 6\times
%10^{-6}\bigg]\ ,
%\end{equation}
%which gives an initial disc mass $M(R,\ t_0)=
%4.8\times 10^{31}\hbox{ g}=0.024M_\odot$.
For a small $x<0.5$, the reduced mass is $m(x)\cong Bx^2/2$
and the evolution of $M(R,\ t)$ may be estimated by
\begin{equation}\label{Mass}
M(R,\ t)\cong aBR^2/(2Gt)\ .
\end{equation}

%If at time $t=t_c$ the radius of circumstellar discs is of order
%$R\sim 1000$ AU and the ionized sound speed $a\sim 20\hbox{
%kms}^{-1}$, it has taken the shock front a timescale of
%$R/(2a)\sim 127\hbox{ yr}$ to reach this distance. This is a
%rather short time and the successive evolution of the disc is well
%described by the downstream (post-shock)
%portions.
%We hence mark the similarity location $x_s=2.00$ with
%the real radius $R$ and characteristic time $t_0$.

%For circumstellar discs of massive stars, we take the radius of an
%ionized disc to be $R\sim 1000$ AU, because discs around
%massive stars are expected to be larger than those around
%T Tauri stars (e.g. Pestalozzi et al. 2004) and they may
%have also expanded somewhat after an IF passes by.
As a self-similar evolution starts from $t\sim 0$, it takes a
timescale of $t_c=R/(2a)\sim 250\hbox{ yr}$ for a shock to reach $R$,
%{\bf what is the $a$ value used here?}
%{\it $a\sim 20\hbox{ km s}^{-1}$ throughout the paper},
where $t_c$ is the calibration time corresponding to a similarity
coordinate $x_c=2$. At this epoch, the disc mass behind a shock is
$M(R,t_c)=a^3t_cm(2)/G\sim0.2M_{\odot}$ where $m(2)=3.4\times 10^{-3}$
by the heavy solid curve of $m(x)$ in Fig. 1. From now on, the shock
continues to travel into distant regions and eventually merges into
ambient H{\sevenrm II} clouds; meanwhile, the subsequent self-similar
evolution of the disc is well described by the downstream (post-shock)
portion. We may then use interpolated values of $m(x)$ with $x\le 2$
%(see Table 2)
and $M(R,\ t)=a^3tm(x)/G$ to estimate $M(R,\ t)$ for $t\ge t_c$.
For a sufficiently small $x$ or large $t$, we may directly use
equation (\ref{Mass}) that shows an enclosed disc mass
proportional to $t^{-1}$. At later times of $5t_c$ ($x=0.4$),
$10t_c$ ($x=0.2$), $20t_c$ ($x=0.1$), $100t_c$ ($x=0.02$),
corresponding to $\sim 1250$ yr, $\sim 2500$ yr, $\sim 5000$ yr,
$\sim 25000$ yr, the enclosed disc mass becomes $\sim 0.03
M_{\odot}$, $\sim 0.015 M_{\odot}$, $\sim 0.0075 M_{\odot}$,
$\sim 0.0015 M_{\odot}$, respectively.

%$0.029 M_{\odot}$, $0.014 M_{\odot}$,$0.007 M_{\odot}$,
%$0.0014 M_{\odot}$

%Now suppose our observed disc is undergoing self-similar evolution
%of the derived form and $M(R,\ t_c)=aBR^2/(2Gt_c)=0.01M_\odot$,
%we get the calibration time $t_c\sim 1.6\times 10^3\hbox{ yr}$.
%The corresponding starting point in the similarity framework is
%$x_c=R/(at_c)\sim 0.15$ so that equation (\ref{Mass}) is valid.

%If the radius of the ionized disc is even larger, say, $R\sim
%10000\hbox{ AU}$ (this is possible since the disc may have
%expanded to this extent at $t=t_c$), the calibration time
%$t_c$ will be $\sim 1.6\times 10^{5}\hbox{ yr}$ if other
%parameters remain unchanged. The corresponding $x_c\sim 0.015$.

Diagnostically, it is not easy to set a criterion for the `disc
disappearance' by decreasing enclosed disc mass $M(R,t)$. We
propose tentatively that a `disc disappearance' corresponds to
a disc mass dropped between one tenth and one hundredth of the
mass at $t=t_c$. This is partly due to the fact that a disc
depletion leads to fainter H$\alpha$ emissions from ionized gas
and infrared/submillimeter emissions from dusts, and partly due
to the possibility that a rarefied disc becomes unstable by
strong stellar wind shears and breaks into clumps
%pieces that may coagulate into nascent planets.
Thus, a timescale of disc disappearance is estimated as
$t_{dis}\sim 10t_c-100t_c$ by expression (\ref{Mass}). For
$t_c\sim 250\hbox{ yr}$, the `disappearance timescale' of
$t_{dis}\sim 2.5\times10^3-2.5\times 10^{4}\hbox{ yr}$ is very
much shorter than the typical timescale of several $\hbox{Myr}$
for low-mass T Tauri stars.
%It is also possible that disc disappearance
%is concurrent with nascent planet formation.

%$B$ is only related to $a$ ?}{\it Yes}.
%
%{\bf If $D$, $a_0$ and $R$ remain the same, but $a$ is one order
%of magnitude larger\footnote{ {\bf This is only to show the trend
%when the ionized sound speed $a$ is even higher. But indeed $a$
%cannot be too high. The highest temperature of ionized gas around
%the most massive stars is $\sim 1.3\times 10^4$ K (Osterbrock
%1989).}} (i.e., $\epsilon$ is one order of magnitude smaller),
%$A\propto \epsilon^2$ will be two orders of magnitude smaller and
%so does $B\propto A$. Numerical procedure shows $m(2)$ will be two
%orders of magnitude smaller too. While the calibration time
%$t_c=R/(2a)$ will be one order of magnitude smaller. These facts
%are coupled together towards a remarkable result that at $t=t_c$
%the disc mass $M(R,t_c)=a^3t_cm(2)/G$ remains unchanged, yet the
%evolution timescale is reduced by one order of magnitude. This may
%apply to even more massive stars which can heat up circumstellar
%discs to even higher temperatures.}

%In addition to the mass depletion rate, we also obtain the
%density and velocity profiles in the disc. At smaller radii,
%the disc tends to have a constant density, while the radial
%expansion velocity becomes linear in radii (see the top and
%middle panels of Fig. 1).

\section{DISCUSSIONS}

While the self-similar process described here is highly idealized,
it may catch several gross features of a disc dispersal process
not fully explored so far. First, we assume a fast traveling IF
together with environmental X-ray and EUV radiation field that
heat up the entire disc to an isothermal state in a short time.
Secondly, rotation and gravity are ignored in the limit of an
overwhelming thermal pressure. Our model of invariant shock is
more applicable to discs around very massive stars or to
circumstellar discs exposed to intense X-ray and EUV radiation
from nearby external sources (e.g., evolved massive stars or
supernova explosions; Chevalier 2000). Thirdly, we ignore the disc
thickness. In reality, such a disc should expand both radially and
vertically besides rotation, with photoevaporation being
concurrent. A thorough analysis of these aspects deems worthwhile.

We now summarize highlights of our model. First, a circumstellar
disc as a whole is expanding rather than fixed or shrinking
shortly after the ignition of nuclear reactions in the stellar
core. This expansion induced disc mass-loss rate is modelled as
self-similar and differs from classic photoevaporation models.
%{\bf The more powerful the central radiative output is (e.g., a
%more massive star), the faster the disc expands and disperses.}
For very massive stars submerged in an intense X-ray and EUV
radiation field, circumstellar discs become too short-lived to be
observed after entering the main-sequence. Secondly, the fast
moving shock
%(i.e., $\sim 2a$) might serve as a trigger of forming protoplanets
serves to disperse disc materials and the rarefied disc may become
vulnerable to fragmentations by stellar wind shears. These processes
accelerate the disc destruction.
%easier and easier to be destroyed by stellar winds, so that planets
%may form very quickly which in return aggravates the disc destruction.
Thirdly, this self-similar expansion tends to establish a more uniform
density distribution even though the initial density profile is
non-uniform. Fourthly, during this self-similar expansion phase, a
considerable mass fraction of the disc is expelled or displaced
to distant places; these processes might explain the presence of
appreciable amounts of ionized gas at intermediate radii (i.e.
$10^3\sim 10^4$ AU) in ultracompact H {\sevenrm II} regions (e.g.,
Wood \& Churchwell 1989; Shu et al. 2002). Moreover, as time goes
on, these dispersed disc materials eventually cool down and may
become a reservoir for producing Edgeworth-Kuiper Belt-like objects
around a central star.

We now comment on several aspects of further model development. First,
even for an isothermal model, the temperatures across a shock can be
allowed to be different (Shen \& Lou 2004b; Bian \& Lou 2005; Yu et al.
2006). Secondly, the isothermal condition can be replaced by the more
general polytropic approximation (Wang \& Lou 2006; Lou \& Gao 2006).
Thirdly, it is of considerable interest to incorporate the effect of
a magnetic field (Shen et al. 2005; Yu \& Lou 2005; Lou \& Zou 2004,
2006; Lou \& Wu 2005; Wu \& Lou 2006; Lou \& Bai 2006).

Finally, we speculate that grossly spherical `champagne flows'
with shocks traveling in H {\sevenrm II} clouds (e.g., Franco et
al. 1990; Shu et al. 2002; Shen \& Lou 2004b; Bian \& Lou 2005) on
much larger spatial scales encompassing a circumstellar disc can
be driven by an intense central stellar radiation, stellar winds
and outflows from the disc sustained by photoevaporation. Triggered
by shocks and various flow or thermal instabilities, fragmentations
would occur in clouds; by cooling and coagulation with time, these
clumps may evolve into comet-like objects. This
might be the origin of the Oort cloud around our solar system
and should be common in other protostellar systems as well.

Not only for young stellar objects, circumstellar discs are also
observed around main-sequence stars or even more evolved stars
(e.g., Zuckerman 2001). These debris dusty discs are exposed
to intense radiation from central objects during the late
evolutionary stage. Disc expansions and shocks can be initiated
in these discs and evolve in a self-similar manner. Even with
considerable idealization, the analysis is technically challenging
as rotation, central gravity and disc self-gravity need to be
taken into account.

\section*{Acknowledgments}
This research was supported in part by the ASCI Center
for Astrophysical Thermonuclear Flashes at the U. of
Chicago under
%DoE
Department of Energy contract B341495, by the
%SFMSBSRP
Special Funds for Major State Basic Science Research Projects
of China, by the THCA, by the Collaborative Research Fund
from the National Natural Science Foundation of China (NSFC)
%NSF of China (NSFC) for YOOCS
Young Outstanding Overseas Chinese Scholars (NSFC 10028306)
at the National Astronomical Observatories,
%NAOC/CAS,
Chinese Academy of Sciences,
by NSFC grants 10373009 and 10533020 at the
Tsinghua Univ., and by the SRFDP 20050003088
%Specialized Research Fund for the Doctoral Program of Higher Education
and the Yangtze Endowment from the Ministry of Education at the Tsinghua
Univ. Affiliated institutions of Y-QL share this contribution.

\end{document}